\documentclass[twocolumn,showpacs,preprintnumbers,amsmath,amssymb]{revtex4}
\usepackage{graphicx}
\usepackage{dcolumn}
\usepackage{bm}
\usepackage{hyperref}

\begin{document}
\title{Parameterizations of Immiscible Two-Phase Flow in Porous Media}
\author{H{\aa}kon Pedersen\email{hakon.pedersen@ntnu.no} and Alex Hansen\email{alex.hansen@ntnu.no}}
\affiliation{PoreLab, Department of Physics, Norwegian University of Science and Technology, 
NO--7491 Trondheim, Norway}
\date{\today {}}
\begin{abstract}
A fundamental variable characterizing immiscible two-phase flow in porous
media is the wetting saturation, which is the ratio between the pore volume
filled with wetting fluid and the total pore volume. More generally, this
variable comes from a specific choice of coordinates on some underlying space,
the domain of variables that can be used to express the volumetric flow rate. The
underlying mathematical structure allows for the introduction of other variables
containing the same information, but which are more convenient from a theoretical
point of view. We introduce along these lines 
polar coordinates on this underlying space, where the angle plays a role similar to the wetting saturation. We derive
relations between these new variables based on the Euler homogeneity theorem.  We
formulate these relations in a coordinate-free fashion using differential forms. Lastly, we discuss
and interpret the co-moving velocity in terms of this coordinate-free representation.      
\end{abstract}
\maketitle
\section{Introduction}
\label{intro}

Flow of immiscible fluids in porous media \cite{b88,s11,b17,ffh22} 
is a problem that has been in the hands of engineers for a long time.  
This has resulted in a schism between the physics at the pore scale 
and the description of the flow at scales where the porous medium 
may be seen as a continuum, also known as the Darcy scale. 
This fact has led to the phenomenological {\it relative permeability\/} 
equations proposed by Wycoff and Botset in 1936 \cite{wb36} 
with the inclusion by Leverett of the concept of capillary pressure in
1940 \cite{l40}, still today being the unchallenged approach to 
calculate flow in porous media at large scales.   
The basic ideas of the theory are easy to grasp. 
Seen from the viewpoint of one of the immiscible fluids, 
the solid matrix and the other fluid together reduce the pore space 
in which that fluid can move. Hence,
the effective permeability as seen by the fluid is reduced, and the permeability 
reduction factor of each fluid is their relative
permeability. The capillary pressure models the interfacial 
tension at the interfaces between the immiscible fluids by
assuming that there is a pressure difference between the pressure field 
in each fluid.  The central variables of the 
theory, the relative permeabilities and the capillary pressure curves, 
are determined routinely in the laboratory under the
name \textit{special core analysis}, and then used as input in reservoir 
simulators \cite{b17,d13}.  A key assumption in the 
theory is that the relative permeabilities and the capillary 
pressure are functions of the saturation alone. This simplifies
the theory tremendously, but it also distances the theory from realism.

Some progress has been made in order to improve on relative permeability
theory.  Barenblatt et al.\ \cite{bps02} recognized that a key assumption 
in relative permeability theory is that the flow is {\it locally\/} in a 
steady state, even if the flow as a whole is developing.  This has as an implication
that the central variables of that theory, the relative permeabilities 
and the capillary pressure, are functions of the saturation alone. They then go on to 
generalize the theory to flow which is locally out of equilibrium, exploring how the
central variables change.  Wang et al.\ \cite{waa19} go further by introducing  
dynamic length scales due to the mixing zone variations, over which the spatial averaging 
is done.     

The relative permeability approach is phenomenological.  Going beyond relative permeability
theory means making a connection between the physics at the pore scale with 
the Darcy scale description of the flow.  The attempts at constructing
a connection between these two scales --- which may stretch from micrometers to
kilometers --- have mainly been focused on homogenization.  That, is 
replacing the original porous medium by an equivalent spatially structure-less one. 

The most famous approach to the scale-up problem along these lines is 
\textit{Thermodynamically Constrained Averaging Theory} (TCAT) 
\cite{hg90,hg93,hg93b,nbh11,gm14}, based on thermodynamically consistent definitions 
made at the continuum scale based on volume averages of pore-scale thermodynamic 
quantities, combined with closure relations based on homogenization \cite{w86}.  
All variables in TCAT are defined in terms of pore-scale variables. However, 
this results in many variables and complicated assumptions are 
needed to derive useful results.

Another homogenization-approach based on non-equilibrium thermodynamics 
uses Euler homogeneity to define the up-scaled pressure.  From this, Kjelstrup et al. derive 
constitutive equations for the flow while keeping the number of variables down 
\cite{kbhhg19,kbhhg19b,bk22}. 

There is also an ongoing effort in constructing a scaled-up theory based on geometrical 
properties of pore space by using Hadwiger theorem \cite{cacbsbgm18,acbrlasb19,cba19}.
The thorem implies that we can express the properties of the spatial geometry of
the three-dimensional porous medium as a 
linear combination of four Minkowski functionals: volume, surface area, mean curvature, and Gaussian curvature.
These are the only four numbers required to characterize the geometric state of
the porous medium. The connectivity of the fluids is described by the Euler
characteristic, which by the Gauss-Bonnet theorem can be computed from the total curvature.

A different class of theories are based on detailed and specific assumptions 
concerning the physics involved. An example is \textit{Local Porosity Theory}
\cite{hb00,h06a,h06b,h06c,hd10,dhh12}.  Another example is the \textit{Decomposition in
Prototype Flow} (DeProf) \textit{theory} which is a fluid mechanical model combined with 
non-equilibrium statistical mechanics based on a classification scheme of 
fluid configurations at the pore level \cite{vcp98,v12,v18}. A third example is
that of Xu and Louge \cite{xl15} who introduce a simple model based on Ising-like statistical 
mechanics to mimic the motion of the immiscible fluids at the 
pore scale, and then scale up by calculating the mean field behavior of the model.
In this way they obtain the wetting fluid retention curve.

The approach we take in this paper is build on the approach to the upscaling problem 
found in References \cite{hsbkgv18,rsh20,rpsh22,hfss22,fsh22}.  The underlying idea is to
map the flow of immiscible fluids in porous media, a dissipative and
 therefore out-of-equilibrium system, onto a system which is in equilibrium.
The \textit{Jaynes principle of maximum entropy} \cite{j57} may then be invoked and the 
scale-up problem is transformed into that of calculating a partition function
\cite{hfss22}.  

In order to sketch this approach, we need to establish the concept of {\it steady-state flow.\/}
In the field of porous media, ``steady-state flow" has two meanings.  The traditional one
is to define it as flow where all fluid interfaces remain static \cite{ea00}. The other one,
which we adopt here, is to state that it is flow where the macroscopic variables remain 
fixed or fluctuate around well-defined averages.  The fluid interfaces will move and on
larger scales than the pore scale one will see fluid clusters breaking up and merging.
If the porous medium is statistically homogeneous, we will see the same local statistics 
describing the fluids everywhere in the porous medium \cite{fsh22}. 

\begin{figure}
\includegraphics[width=10truecm]{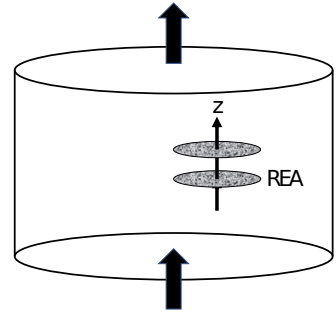}
\caption{A porous plug in the form of a cylinder.  There is a mixture of two immiscible fluids flowing
through it in the direction of the cylinder axis under steady-state conditions.  We furthermore imagine two
disks orthogonal to the cylinder axis.  These are REAs. }
\label{fig1}
\end{figure}

Imagine a porous plug as shown in Figure \ref{fig1}. There is a mixture of two immiscible fluids 
flowing through it in the direction of the cylinder axis under steady-state conditions. 

A central concept in the following is the \textit{Representative Elementary Area} (REA) \cite{bb12,rpsh22,hfss22}. 
We pick a point in the porous plug. In a neighbourhood of this point, there will a set of streamlines associated
with the velocity field $\vec v_p$. The overall flow direction
is then defined by the tangent vectors to the stramlines. We place an imaginary disk of area $A$ at
the point, with orientation such that $A$ is orthogonal to the overall flow direction. We illustrate this as
the lower disk in Figure \ref{fig1}.  We assume that the disk is small enough for 
the porous medium to be homogeneous over the size of the plane with respect to porosity and 
permeability. This disk is an REA.

The REA will contain different fluid configurations.  By ``fluid
configuration'', we mean the spatial distributions of the two fluids in the disk and their scalar velocity fields.  We also show a second disk in Figure
\ref{fig1}, which represents another REA placed further along the average flow.
Since the flow is incompressible, we can imagine the $z$-axis as corresponding
to a pseudo-time axis. We may state that
the fluid configuration in the lower disk of  Figure \ref{fig1} evolves into the
fluid configuration in the upper disk under pseudo-time-translation. We now associate an entropy to the fluid configurations in the sense of Shannon \cite{s48}.  Even though the system is producing molecular
entropy through dissipation, it is \textit{not} producing Shannon entropy along the $z$-axis. This allows us to
use the Jaynes maximum entropy principle, and a statistical mechanics based on Shannon entropy ensues
\cite{hfss22}. 

This statistical mechanics scales up the pore-scale physics, represented by the fluid configurations to
the Darcy scale, which then is represented by a thermodynamics-like formalism involving averaged 
velocities.

Rudiments of this thermodynamics-like formalism was first studied by   
Hansen et al.\ \cite{hsbkgv18}, who used extensivity to derive a set of equations that relate the 
seepage velocity of each of two immiscible fluids flowing under steady state 
conditions through a representative volume element (REV). They introduced a 
{\it co-moving velocity\/} that together with the average seepage velocity $v$ contained
the necessary information to determine the seepage velocities of the more wetting and the 
less wetting fluids, $v_w$ and $v_n$ respectively.  Stated in a different way, these equations  
made it possible to make the transformation
\begin{equation}
  \label{eq:main-map}
  (v,v_m) \leftrightarrows (v_w,v_n) \ \;,
\end{equation}
where $S_w$ and $S_n$ are the wetting and non-wetting saturations respectively. 
    
It is the aim of this paper to present a {\it geometrical interpretation\/} of the transformation
in equation \ref{eq:main-map}. We wish to provide
an {\it intuitive\/} understanding of precisely what the transformation is doing and
determine the role of the co-moving velocity.  Our aim is not to develop the theory presented in
\cite{hsbkgv18} further by including new results, but rather consolidate and amend the existing
theory with a deeper understanding.

Our geometrical interpretation will rest upon the definition of a space spanned by two central
variables: {\it the wetting and non-wetting transversal pore areas.\/}  These areas are defined as the area
of REAs covered by the wetting and non-wetting fluids respectively.   

Instead of using the two extensive areas directly, one often work with coordinates where one is the (wetting) 
saturation. If the pore area of the REA is fixed, it is often convenient to let
the other variable be this area. A third way --- which is new --- is to use
{\it polar coordinates.\/} This, as we shall see, simplify the theory considerably. 

In Section \ref{rea} we describe the REA and the relevant variables.  We define 
intensive and extensive variables, tying them to how they scale under scaling of the size of 
the REV.   We also review the central results of Hansen et al.\
\cite{hsbkgv18} here: the introduction of the auxiliary {\it thermodynamic velocities\/} and the 
{\it co-moving velocity.\/}  

The central concept in our approach to the immiscible two-fluid flow problem is
that of the pore areas. In Section \ref{area} we introduce {\it area space\/}, a
space that simplifies the analysis considerably. We showcase different
coordinate systems on this space, focusing on  polar coordinates.

Section \ref{euler} expresses the central results of Hansen et al.\ \cite{hsbkgv18} in polar
coordinates. Expressed in this coordinate system, the geometrical structure implied by the
extensivity versus intensivity conditions impose restrictions that simpliefies
the equations.  The section goes on to express
the total pore velocitiy, seepage velocities, thermodynamic velocities and co-moving velocity
in terms of polar coordinates.  This provides a clear geometrical interpretation of the 
co-moving velocity.

Section \ref{coordinate} focus on relations between the different velocities
that can be expressed without referring to any coordinate system.  We do this by invoking
{\it differential forms\/} and {\it exterior algebra\/} \cite{f63,mtw17}.  Differential
forms constitute the generalization of the infinitesimal line, area or volume elements used in 
integrals and exterior calculus is the algebra that makes it possible to handle them. Flanders
predicted in the early sixties \cite{f63} that within short they would become important tools 
in engineering. This did not happen, and they have remained primarily within
mathematics and theoretical physics. Such objects are also especially prevalent in
thermodynamics, and we will here show that the computational
rules of differential forms provide a simple way of expressing the relations
between the velocities in the system at hand.

Section \ref{comov} focuses the discussion on the co-moving velocity by
constructing a coordinate-independent expression
that defines it.                     
       
In the last Section \ref{conclusion}, we summarize our main results which may be stated through three equations,
(\ref{eq18}), (\ref{eq29-1}) and (\ref{eqn53}). These equations formulate in a coordinate-free way
the three relations that exist between the average seepage velocity and the co-moving velocity. 

\section{Representative Elementary Area}
\label{rea}

We will now elaborate on the definition of the Representative Elementary Area
(REA) presented in Section \ref{intro}.

We define the {\it transversal pore area\/} $A_p$ as previously, which defines a porosity
\begin{equation}
\label{e5}
\phi=\frac{A_p}{A}\;.
\end{equation}

The pore volume contains two immiscible fluids; the more wetting fluid (to be referred to as the 
\textit{wetting fluid}) or the less wetting fluid (to be referred to as the \textit{non-wetting} fluid).  The transversal
pore area $A_p$ may therefore be split into the area of the REA disk cutting through wetting
fluid $A_w$, or cutting through the non-wetting fluid $A_n$, so that we have
\begin{equation}
\label{eq1}
A_p=A_w+A_n\;.
\end{equation}
We also define the saturations
\begin{eqnarray}
S_w=\frac{A_w}{A_p}\;,\label{eq2}\\
S_n=\frac{A_n}{A_p}\;,\label{e13}\\
\nonumber
\end{eqnarray}
so that
\begin{equation}  
\label{e10}
S_w+S_n=1\;.
\end{equation}

We note that the transversal pore areas $A_p$, $A_w$ and $A_n$ are {\it extensive\/}
in the REA area $A$. That is
$A_p\to\lambda A_p$, $A_w\to\lambda A_w$ and $A_n\to\lambda A_n$ when $A\to\lambda A$. 
The porosity $\phi$ and the two saturations $S_w$ and $S_n$ are
{\it intensive\/} in the REA area $A$: $\phi\to\lambda^0\phi$, $S_w\to\lambda^0 S_w$
and $S_n\to\lambda^0 S_n$ when $A\to\lambda A$.

There is a time averaged volumetric flow rate $Q$ through the REA. The volumetric
flow rate consists of two components, $Q_{w}$ and $Q_{n}$, which are the
volumetric flow rates of the wetting and the non-wetting fluids. We have 
\begin{equation}
\label{e6}
Q=Q_{w}+Q_{n}\;.  
\end{equation}
We define the total, wetting and non-wetting {\it seepage velocities\/} respectively as
\begin{eqnarray}
v=\frac{Q}{A_p}\;,\label{e7}\\
v_w=\frac{Q_w}{A_w}\;,\label{e8}\\
v_n=\frac{Q_n}{A_n}\;.\label{e9}
\end{eqnarray}

Using equations (\ref{e6}) to (\ref{e9}), we find
\begin{eqnarray}
v&=&\frac{Q}{A_p}=\frac{A_w}{A_p}\ \frac{Q_w}{A_w}+\frac{A_n}{A_p}\ \frac{Q_n}{A_n}\nonumber\\
 &=&S_w v_w+S_n v_n\;.\label{e14}
\end{eqnarray}

The volumetric flow rates $Q$, $Q_w$ and $Q_n$ are extensive in the REA area $A$
and the velocities $v$, $v_w$ and $v_n$ are intensive in the area $A$. 

\subsection{Euler homogeneity}
\label{deform}

As the areas $A_w$ and $A_n$ and the volumetric flow rate $Q$ are
extensive in the REA area $A$, we have that
\begin{equation}  
\label{eq7}
Q(\lambda A_w,\lambda A_n)=\lambda Q(A_w,A_n)\;.
\end{equation} 
This scaling was the basis for the theory presented in \cite{hsbkgv18},
which we now summarize. 
  
We now apply the \textit{Euler homogeneous function theorem} to $Q$. We take the derivative with respect to $\lambda $ on both sides of 
(\ref{eq7}) and set $\lambda =1$. This gives
\begin{equation}
\label{eqn11}
Q(A_{w},A_{n})=A_{w}\left( \frac{\partial Q}{\partial A_{w}}\right)
_{A_{n}}+A_{n}\left( \frac{\partial Q}{\partial A_{n}}\right)
_{A_{w}}\;.
\end{equation}
We note here that we assume that $A_w$ and $A_n$ are our independent
control variables.  This makes $A_p$, $S_w$ and $S_n$ as dependent variables.
More precisely, equation (\ref{eqn11}) is 
the Euler theorem for homogeneous functions applied to a degree-$1$ homogeneous
function $Q$. By dividing this
equation by $A_p$, we have 
\begin{equation}
\label{eqn10-1}
v=S_{w}\left( \frac{\partial Q}{\partial A_w}\right)
_{A_n}+S_{n}\left( \frac{\partial Q}{\partial A_n}\right)_{A_w}\;,  
\end{equation}
where we have used equations (\ref{eq2}) and (\ref{e13}).
We define the two {\it thermodynamic velocities\/} $\hat{v}_w$ and $\hat{v}_n$ as
\begin{equation}
\label{eqn10-2}
\hat{v}_{w}=\left( \frac{\partial Q}{\partial A_{w}}\right)_{A_{n}}\;,  
\end{equation}
and
\begin{equation}
\label{eqn10-3}
\hat{v}_{n}=\left( \frac{\partial Q}{\partial A_{n}}\right)_{A_{w}}\;,  
\end{equation}
so that we may write (\ref{eqn10-1}) as
\begin{equation}
\label{eqn10-5}
v=S_w \hat{v}_w+S_n\hat{v}_n\;.
\end{equation}

The thermodynamic velocities $\hat{v}_w$ and $\hat{v}_n$ are {\it not\/} the physical 
velocities $v_w$ and $v_n$.  Rather, the {\it most general relation\/} 
between $v_w$ and $\hat{v}_w$, and $v_n$ and $\hat{v}_n$, that fulfills both 
equations (\ref{e14}) and (\ref{eqn10-5}),
\begin{equation}
v=S_w\hat{v}_w+S_n\hat{v}_n=S_wv_w+S_nv_n\;,
\end{equation} 
can be expressed as
\begin{align}
  \hat{v}_{w} \ =& \ v_{w}+S_n v_m \label{eqn10-5.5} \\
  \hat{v}_{n}\ =& \ v_{n}-S_w v_m \label{eqn10-5.6} \;.
\end{align} 
This \textit{defines} the {\it co-moving velocity,\/} $v_m$, which  
relate the thermodynamic and the physical velocities.

We have up to now used $(A_w,A_n)$ as our control variables.  If we now consider
the coordinates $(S_w,A_p)$ instead,  we can write
\begin{eqnarray}
\label{eq1000}
\hat{v}_{w}=\left( \frac{\partial Q}{\partial A_{w}}\right)_{A_{n}}
&=&\left(\frac{\partial Q}{\partial S_w}\right)_{A_p}\left(\frac{\partial S_w}{\partial A_w}\right)_{A_n}
+\left(\frac{\partial Q}{\partial A_p}\right)_{S_w}\nonumber\\
&=& \left( \frac{d v}{d S_w} \right)S_n+v\;,
\end{eqnarray}
where we have used that
\begin{equation}
\label{eq1001}
\left(\frac{\partial S_w}{\partial A_w}\right)_{A_n}=\left(\frac{\partial}{\partial A_w}\right)_{A_n}\left(\frac{A_w}{A_w+A_n}\right)
=\frac{S_n}{A_p}\;.
\end{equation}
Likewise, we find that
\begin{equation}
\label{eq1002}
\hat{v}_{n}=\left( \frac{\partial Q}{\partial A_{n}}\right)_{A_{w}}=- \left( \frac{d v}{ d S_w}\ \right) S_w+v\;.
\end{equation}

We combine equations (\ref{eq1000}) and (\ref{eq1002}) with equations (\ref{eqn10-5.5}) and (\ref{eqn10-5.6}) to find
\begin{align}
v_w \ =& \ v+S_n\left(\frac{dv}{dS_w}-v_m\right)\;,\label{e1} \\
v_n \ =& \ v-S_w\left(\frac{dv}{dS_w}-v_m\right)\;,\label{e2}
\end{align}
We see that these two equations gives the map $(v,v_m) \rightarrow
(v_w,v_n)$. 

We now subtract equation (\ref{e2}) from (\ref{e1}), finding
\begin{equation}
\label{eq1003}
\frac{dv}{dS_w}=v_w-v_n+v_m\;.
\end{equation}
We now differentiate equation (\ref{e14}) with respect to $S_w$,
\begin{equation}
\label{eq1004}
\frac{dv}{dS_w}=v_w-v_n+S_w\frac{dv_w}{dS_w}+S_n\frac{dv_n}{dS_w}\;.
\end{equation}
We compare equations (\ref{eq1003}) and (\ref{eq1004}), finding
\begin{equation}
\label{eq1005}
v_m=S_w\frac{dv_w}{dS_w}+S_n\frac{dv_n}{dS_w}\;.
\end{equation}
Equations (\ref{e14}) and (\ref{eq1005}) constitute the inverse mapping $(v_w,v_n)\to (v,v_m)$. 

The mapping $(v,v_m)\to (v_w,v_n)$ tells us that given the constitutive equations for 
$v$ and $v_m$, we also have the constitutive equations for $v_w$ and $v_n$.  It turns out that
the constitutive equation for $v_m$ is surprisingly simple \cite{rpsh22}
\begin{equation}
\label{eq1006}
v_m=a+b\frac{dv}{dS_w}\;,
\end{equation}
where $a$ and $b$ are coefficients depending on the entropy associated with the fluid configurations
\cite{hfss22}.    

The equations we have used in Section \ref{rea} hint at an underlying mathematical
structure which may seem complex.  As we will now show, it is in fact quite the opposite. 

\section{Coordinate Systems in Area Space}       
\label{area}

\begin{figure}
\includegraphics[width=6truecm]{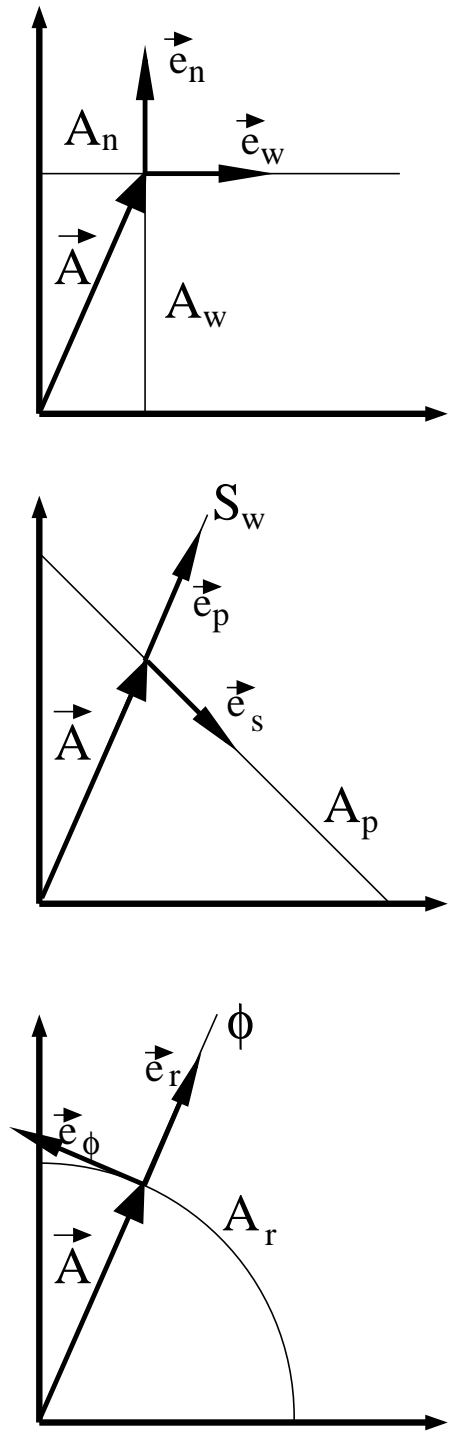}
\caption{We illustrate the three coordinate systems we use to parametrize the 
transversal pore area.  The {\it top figure\/} shows the Cartesian coordinate system $(A_w,A_n)$. 
A curve of constant  $A_w$ and $A_n$ is indicated as $A_w$ and $A_n$ respectively. Both $A_w$ and $A_n$ are extensive.  
The orthonormal basis set $(\vec e_w,\vec e_n)$ is shown as attached to the
point $\left( A_w , A_n \right)$. We show in the {\it middle figure\/}
 the saturation coordinate system $\left( A_p, S_w \right)$.  We indicate a
 curve of constant $A_p$ and $S_w$ as $A_p$ and
$S_w$ in the figure. $A_p$ is extensive and $S_w$ is intensive.
We show the basis set $(\vec e_p,\vec e_s)$, which is not orthonormal. In the {\it lower figure\/}
we show the polar coordinate system, with curves of constant $A_r$ and $\phi$ denoted $A_r$ and $\phi$ respectively.  $A_r$ is extensive and $\phi$ is intensive. The 
basis set $(\vec e_r,\vec e_\phi)$ is orthonormal. In all three figures, we show the transversal
pore area as a vector $\vec A$.} 
\label{fig2}
\end{figure}

The transversal pore areas $A_w \ge 0$ and $A_n \ge 0$ parametrize the first 
quadrant of $\mathbb{R}^{2}$. This serves as the ``area space'' mentioned
earlier,which we will continue to denote as such. It is important to realize that this space is {\it not\/} 
physical space. A value of the transversal pore area $A_{p}$ corresponds to a point
$\left( A_w, A_n \right)$ in this space. If we consider the entire plane $\mathbb{R}^{2}$ as a whole
and simply restrict our attention to the first quadrant, we may treat the area
space as a vector space. The point $\left( A_w, A_n \right)$ may then
equivalently be described by a vector,
\begin{equation}
\label{e100}
\vec A=A_w \vec e_w+A_n\vec e_n\;,
\end{equation}
where $\vec e_w$ and $\vec e_n$ form an orthonormal basis set,
see the upper figure in Figure \ref{fig2}. Note that in this picture, the bases
are shown as attached to the point $\left(A_w, A_n \right)$. We reiterate that both $A_w$ and $A_n$ are extensive.   

Every point in the transversal area space corresponds to a a given saturation
$S_w$ and a transversal pore area $A_p$ and we may view the map 
$(A_w,A_n)\to(A_p,S_w)$  given by equation (\ref{eq1}) and (\ref{eq2}) as a coordinate transformation.
This is a more natural coordinate system to work with since $S_w$ is an intensive variable and $A_p$ is
in practice kept constant.      
We name this system the {\it saturation coordinate system.\/}  We show the normalized basis vector 
set in the middle figure in Figure \ref{fig2}.  We calculate 
\begin{eqnarray}
\vec u_p&=&\left(\frac{\partial \vec A}{\partial A_p}\right)_{S_w}= S_w \vec e_w+(1-S_w)\vec e_n\;,\label{e101}\\
\vec u_s&=&\left(\frac{\partial \vec A}{\partial S_w}\right)_{A_p}= A_p \vec e_w-A_p\vec e_n\;.\label{e102}
\end{eqnarray}
We normalize the two vectors $\vec u_p$ and $\vec u_s$ to find
\begin{eqnarray}
\vec e_p&=&\frac{S_w\vec e_w+(1-S_w)\vec e_n}{[1-2S_w(1-S_w)]^{1/2}}\;,\label{e103}\\
\vec e_s&=& \frac{\vec e_w-\vec e_n}{\sqrt{2}}\;.\label{e104}
\end{eqnarray}
This basis set is not orthonormal, as illustrated in Figure \ref{fig2}.  By expressing $\vec A$ in
this coordinate system, we find
\begin{equation}
\label{e104-1}
\vec A=A_p \left[1-2S_w(1-S_w)\right]^{1/2} \vec e_p\;.
\end{equation}

As will become apparent, the most convenient coordinate system to work with from
a theoretical point of view is  {\it polar coordinates\/} $\left(A_r, \phi
\right)$, given by 
\begin{align}
  A_r \ =& \ \sqrt{A_w^2+A_n^2}\;,  \label{eq105} \\
  \phi \ =& \ \arctan\left(\frac{A_n}{A_w}\right)\;, \label{eq106} 
\end{align} 
or {\it vice versa\/}
\begin{align}
  A_w \ = \ A_r \cos\phi\;,  \label{eq107} \\
  A_n \ = \ A_r \sin\phi\;. \label{eq108}
\end{align} 
We note that $\phi$ is intensive and $A_r$ is extensive. We show the polar
coordinate system in the lower figure in Figure\ \ref{fig2}.

The basis vector set $(\vec e_r,\vec e_\phi)$ is orthonormal, where
\begin{align}
  {\vec e}_r \ =& \ \cos\phi\ {\vec e}_w+\sin\phi\ {\vec e}_n=\vec e_p\;,   \label{e108-1} \\
{\vec e}_\phi \ =& \ -\sin\phi\ {\vec e}_w+\cos\phi\ {\vec e}_n\;.  \label{e108-2}
\end{align}
As for the saturation coordinate system, there is one 
intensive variable, $\phi$, and one extensive variable, $A_r$.
However, in contrast to the saturation coordinate system, both variables
are varied in practical situations.     

We have that 
\begin{equation}
\label{e106}
\vec A=A_r\vec e_r\;.
\end{equation}

We see that this is consistent with equation (\ref{e104-1}) since 
\begin{equation}
\label{e106-1}
A_r=A_p\left[1-2S_w(1-S_w)\right]^{1/2}\;
\end{equation}
and
$\vec e_p=\vec e_r$.

\section{Euler Homogeneity in Polar Coordinates}      
\label{euler}

We now focus on equation (\ref{eq7}) which we repeat here, 
\begin{equation}
Q(\lambda A_w, \lambda A_n)=\lambda Q(A_w,A_n)\;.\nonumber
\end{equation} 
We may interpret this equation geometrically. If we follow the value 
of $Q$ along a ray passing through the origin 
of the two-dimensional space spanned by $(A_w,A_n)$ 
keeping the ratio $A_n/A_w$ constant, it grows linearly
with the distance from the origin.  

In {\it polar coordinates,\/} this means
\begin{equation}
\label{eq8}
Q(A_r,\phi)=\hat v_r A_r\;,
\end{equation}
where
\begin{equation}
\label{eq9}
\hat v_r=\left(\frac{\partial Q}{\partial A_r}\right)_{\phi} =\hat v_r(\phi)\;.
\end{equation}
The important point here is that $\hat v_r$ is {\it not\/} a function of $A_r$.

\begin{figure}
\includegraphics[width=8truecm]{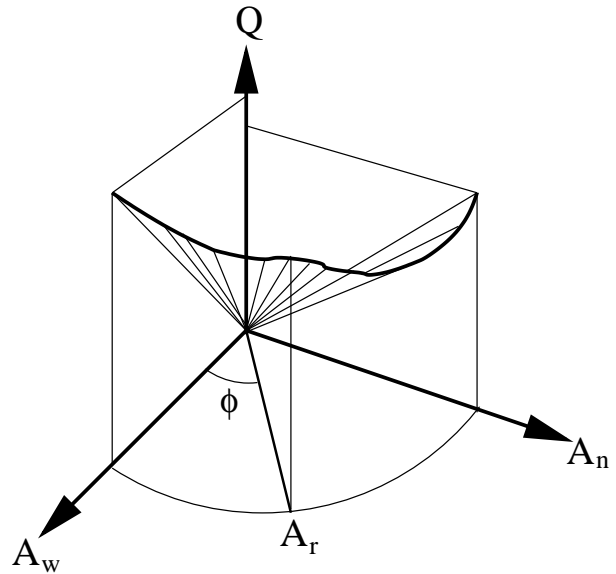}
\caption{We illustrate here the geometrical meaning of $Q$ being Euler homogeneous, 
i.e., fulfilling equation (\ref{eq7}):
Moving along a ray in the space spanned by the $Q$, $A_w$ and $A_n$ directions, 
$Q$ increases linearly with the 
distance from the origin $(Q,A_w,A_n)=(0,0,0)$.  This demonstrates that $Q$ appears as a 
``crumpled" cone originating at the origin.   
Hence, polar coordinates $(A_r,\phi)$ are the most convenient to use 
compared to either the $(A_w,A_n)$ or $(A_p,S_w)$ coordinate 
systems.} 
\label{fig3}
\end{figure}

We may derive equation (\ref{eq8}) from equation (\ref{eq7}) as follows.  First, write
$A_w$ and $A_n$ in terms of $A_r$ and $\phi$ in equation (\ref{eq7}) so that
\begin{eqnarray}
\label{eq8-1}
&&\frac{1}{\lambda}\ Q(\lambda A_r\cos\phi,\lambda A_r\sin\phi)\nonumber\\
&=&Q(A_r\cos\phi,A_r\sin\phi)=Q(A_r,\phi)\;.
\end{eqnarray}
Next, set $\lambda=1/A_r$ to find
\begin{equation}
\label{eq8-2}
A_r\ Q(\cos\phi,\sin\phi)=A_r\hat v_r(\phi)=Q(A_r,\phi)\;,
\end{equation}
where
\begin{equation}
\label{eq8-3-1}
\hat v_r(\phi) \equiv Q(\cos\phi,\sin\phi)
\end{equation}
--- and we are done. We illustrate in Figure \ref{fig3} the geometrical 
meaning of equation (\ref{eq7}), i.e., Euler homogeneity.  The graph of the volumetric flow
rate $Q$ takes the shape of a ``crumpled" cone in the space spanned by $(Q,A_w,A_n)$. 

We now express the different fluid velocities, namely the pore velocity, the seepage velocities,
the thermodynamic velocities and the co-moving velocity, in terms of polar coordinates.

\subsection{Seepage velocities}      
\label{velocities}

The area space spanned by $\left( A_w, A_n \right)$ is simply (a subset of) the
real plane. We can therefore identify it
with the \textit{tangent space} of velocities, which is the same Cartesian plane
as that of the areas $\left( A_w, A_n \right)$. We can therefore regard the seepage velocity as a vector in the area space,
\begin{equation}
\label{eq10}
\vec v=v_w \vec e_w+v_n\vec e_n\;.
\end{equation}
The volumetric flow rate $Q$ is then given by the scalar product
\begin{equation}
\label{eq11}
Q=\vec v \cdot \vec A=v_wA_w+v_nA_n\;.
\end{equation}
In polar coordinates, the seepage velocity is given by 
\begin{equation}
\label{eq14-1}
\vec v=v_r \vec e_r+v_\phi \vec e_\phi\;.
\end{equation}
Hence, the volumetric flow rate in polar coordinates is given by
\begin{equation}
\label{eq12}
Q=\vec v \cdot \vec A=(v_r\vec e_r+v_\phi\vec e_\phi)\cdot(A_r\vec e_r)=v_r A_r\;.
\end{equation}
Comparing with equation (\ref{eq8}), we find
\begin{equation}
\label{eq13}
v_r=\hat v_r=\hat v_r(\phi)\;.
\end{equation}
By using equations (\ref{eq10}) and (\ref{eq14-1}) combined with equations (\ref{e108-1}) and (\ref{e108-2}), we find
\begin{eqnarray}
\label{eq14}
v_r&=&v_w\cos\phi+v_n\sin\phi\nonumber\\
&=&\frac{S_w v_w+(1-S_w)v_n}{[1-2S_w(1-S_w)]^{1/2}}\;,\\
v_\phi&=&-v_w\sin\phi+v_n\cos\phi\nonumber\\
&=&\frac{-(1-S_w)v_w+S_wv_n}{[1-2S_w(1-S_w)]^{1/2}}\;.
\end{eqnarray}
We have here expressed $\cos\phi$ and $\sin\phi$ in terms of $S_w$. 

We now compare equation (\ref{eq8}) with equation (\ref{e7}), rewritten as
\begin{equation}
\label{eq8-3}
Q=vA_p\;.
\end{equation}
Using equation (\ref{e106-1}), we find that their equality demands
\begin{equation}
\label{eq8-4}
v=\frac{\hat v_r}{\left[1-2S_w(1-S_w)\right]^{1/2}}\;.
\end{equation}
Hence, $v$ is \textit{not} the norm of $\vec v$.
   
\subsection{Thermodynamic velocities}      
\label{thermodynamic}

We may write equation (\ref{eqn11}) as
\begin{equation}
\label{eq18}
Q=\hat v_w A_w+\hat v_n A_n = \hat{\vec v} \cdot \vec A\;,
\end{equation}
where we have defined, in the same manner as for equation (\ref{eq10}),
\begin{equation}
\label{eq19}
\hat{\vec v}=\hat v_w \vec e_w+\hat v_n\vec e_n\;.
\end{equation}
We use equations (\ref{e108-1}) and (\ref{e108-2}) to express this equation in
polar coordinates,
\begin{equation}
\label{eq20-1}
\hat{\vec v}=\hat v_r \vec e_r+\hat v_\phi\vec e_\phi\;,
\end{equation}
where
\begin{eqnarray}
\hat v_r&=&\hat v_w\cos\phi+\hat v_n\sin\phi\;,\label{eq20-2}\\
\hat v_\phi&=&-\hat v_w\sin\phi+\hat v_n\cos\phi\;.\label{eq20-3}
\end{eqnarray}

\subsection{Co-moving velocity}      
\label{co-moving}

We defined the co-moving velocity $v_m$ as the velocity function that relates the 
physical seepage velocities and the thermodynamic velocities, see equations 
(\ref{eqn10-5.5}) and (\ref{eqn10-5.6}).  We use these two equations together with
equations (\ref{eq10}) and (\ref{eq19}) to form the vector
\begin{eqnarray}
\label{eq23}
\hat{\vec v}-\vec v&=&(\hat v_w-v_w)\vec e_w+(\hat v_n-v_n)\vec e_n\nonumber\\
&=&\frac{v_m}{A_p}\left(A_n\vec e_w-A_w\vec e_n\right)\nonumber\\
&=&\frac{v_m}{A_p}\ \left(A_r\sin\phi\ \vec e_w-A_r\cos\phi\ \vec e_n\right)\;,
\end{eqnarray}
where we have used that $S_w=A_w/A_p$ and equations (\ref{eq107}) and (\ref{eq108}).
We rewrite this equation in terms of $\vec e_\phi$, equation (\ref{e108-2}), finding
\begin{equation}
\label{eq23-2}
\hat{\vec v}-\vec v=-\frac{v_m}{\cos\phi+\sin\phi}\ \vec e_\phi\;.
\end{equation}
We define
\begin{equation}
\label{eq26}
\hat v_m=\frac{v_m}{\cos\phi+\sin\phi}
\end{equation}
so that equation (\ref{eq23}) may be written
\begin{equation}
\label{eq27}
\vec v=\hat{\vec v}+\hat{\vec v}_m\;,
\end{equation}
where
\begin{equation}
\label{eq28}
\hat{\vec v}_m=\hat v_m\vec e_\phi\;.
\end{equation}

\begin{figure}
\includegraphics[width=8truecm]{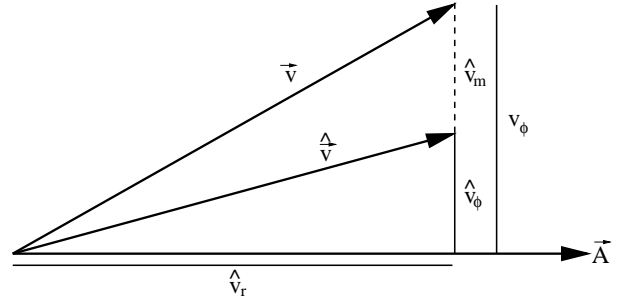}
\caption{We illustrate here equations (\ref{eq12}) and (\ref{eq18}), 
  $Q=\vec v \cdot \vec A=\hat{\vec v}\cdot \vec A =\hat v_r A_r$, where $A_r=|\vec A|$.
  The difference in the vectors $\hat{\vec v}$ and $\vec v$, $\hat{v}_m$, is orthogonal to $\vec A$,  
and cannot be detected by properties of $\vec A$ alone.}
\label{fig4}
\end{figure}

In polar coordinates, the relation in equation (\ref{eq27}) is isolated to the
$\phi$-coordinate. Using equations (\ref{eq14-1}) and (\ref{eq20-3}), we can
express the relation between the vector components $\hat{v}_{\phi}$ and $v_{\phi}$ as
\begin{equation}
\label{eq29}
v_\phi=\hat v_\phi+\hat v_m\;.
\end{equation}
We illustrate this relation in Figure \ref{fig4}.  This figure also demonstrates why
$Q=\hat{\vec v}\cdot\vec A={\vec v}\cdot \vec A$; it is due to $({\vec v}-\hat{\vec v} )=\hat{\vec v}_m\perp
\vec A$. Hence, we have that
\begin{equation}
\label{eq29-1}
\hat {\vec v}_m\cdot \vec A=0\;.
\end{equation}

\section{Coordinate-Free Representation}      
\label{coordinate}

Since the difference between the thermodynamic- and seepage velocities in the
previous section was shown to sit in the \textit{tangent vector components}, we can
benefit from a examining the problem from a position where these components are
easier to work with. In this section, we therefore introduce {\it differential
  forms\/} and {\it exterior algebra\/} \cite{f63,mtw17} to make this treatment
more economic. In addition, this makes it easier to formulate the equations met
so far in a way that does not depend on the coordinate system used.

Equation (\ref{eq8}), $Q=\hat v_r A_r$, may be differentiated to give
\begin{equation}
\label{eqn1000}
dQ=\hat v_r dA_r+d\hat v_r A_r=\hat v_r dA_r+ \hat v_r'A_r d\phi\;,
\end{equation}
where $\hat v_r'=d\hat v_r/d\phi$. Here, $dQ$, $dA_r$ and $d\phi$ are 
one-forms. We furthermore have that
\begin{eqnarray}
\label{eqn1002}
dQ&=&\left(\frac{\partial Q}{\partial A_r}\right)_{\phi}dA_r+
\left(\frac{\partial Q}{\partial \phi}\right)_{A_r}d\phi\nonumber\\
&=&\hat v_rdA_r+\hat v_\phi A_rd\phi\;.
\end{eqnarray}
Comparing equations (\ref{eqn1000}) and (\ref{eqn1002}) gives
\begin{equation}
\label{eqn1003}
\hat v_\phi=\hat v_r'\;.
\end{equation}

We do the same using the coordinate system $(A_w,A_n)$.  From equation
(\ref{eq11}), we find
\begin{eqnarray}
\label{eqn1004}
dQ&=&\hat v_w dA_w+\hat v_n dA_n+A_w d\hat v_w+A_nd \hat v_n\nonumber\\
&=&\hat v_w dA_w+\hat v_n dA_n\;.
\end{eqnarray}
This implies the relation
\begin{equation}
\label{eqn1005}
A_w d\hat v_w+A_n d\hat v_n=0\;.
\end{equation}
This equation corresponds to the {\it Gibbs-Duhem\/} equation in thermodynamics,
which is a statement of the dependency amongst the intensive variables.

We express $\hat v_w$ and $\hat v_n$ in terms of $\hat v_r$ and $\hat v_\phi$ by
inverting equations (\ref{eq20-2}) and (\ref{eq20-3}) to find
\begin{align}
  \hat v_w \ = \ \hat v_r\cos\phi-\hat v_\phi\sin\phi\;, \label{eq31} \\
  \hat v_n \ = \ \hat v_r\sin\phi+\hat v_\phi\cos\phi\;.  \label{eq32}
\end{align}
By applying the exterior derivative, we get the relations
\begin{align}
  d\hat v_w \ = \ \cos\phi\ d\hat v_r-\sin\phi\ d\hat v_\phi-\hat v_n d\phi\;, \label{eq33} \\
  d\hat v_n \ = \ \sin\phi\ d\hat v_r+\cos\phi\ d\hat v_\phi+\hat v_w d\phi\;. \label{eq34}
\end{align}
Combining equations (\ref{eq31}) to (\ref{eq34}) with the Gibbs-Duhem equation
(\ref{eqn1005}) gives
\begin{equation}
\label{eqn1006}
\frac{d\hat v_r}{d\phi}=\hat v_\phi\;,
\end{equation}
which is identical to equation (\ref{eqn1003}).  Hence, this is the
Gibbs-Duhem equation in polar coordinates. 

The thermodynamic velocity, equation (\ref{eq19}), is therefore given by
\begin{equation}
\label{eq37}
\hat{\vec v}=\hat v_r\vec e_r + \hat v_r^{\prime}\vec e_\phi\;.
\end{equation}

We now have one of the main results of Hansen et al.\ \cite{hsbkgv18} in a very compact form,
\begin{equation}
\label{eq38}
\vec v=\hat v_r\vec e_r+(\hat v_r'+\hat v_m)\vec e_\phi\;.
\end{equation}
We note that since $\hat v_r$ only depends on $\phi$, the same must be true for $\hat v_m$.  Hence,
$\hat v_m=\hat v_m(\phi)$. We here see that the relation between the thermodynamic-
and seepage velocities are not captured in the extensive structure tied to the
radial coordinate, but instead in the intensive quantities which only has a
dependency on $\phi$. \\

We now differentiate equation (\ref{eq8-3}), $Q={\vec v}\cdot{\vec A}$ written in the saturation coordinate
system, to find
\begin{equation}
\label{eq39}
dQ=v\ dA_p+\frac{dv}{dS_w} A_pdS_w\;.
\end{equation}
We note that 
\begin{equation}
\label{eq40}
dA_p=dA_w+dA_n\;,
\end{equation}
from differentiating equation (\ref{eq1}).  Likewise, we note that
\begin{eqnarray}
\label{eq41}
dS_w&=&\left(\frac{\partial S_w}{\partial A_w}\right)_{A_n}dA_w
+\left(\frac{\partial S_w}{\partial A_n}\right)_{A_w}dA_n\nonumber\\
&=&\frac{1}{A_p^2}\left(A_n dA_w-A_w dA_n\right)\;,
\end{eqnarray}
using equation (\ref{eq2}). We combine equations (\ref{eq39}) to
(\ref{eq41}) and find
\begin{equation}
\label{eq42}
dQ=\left(v+(1-S_w)\frac{dv}{dS_w}\right)dA_w+\left(v-S_w\frac{dv}{dS_w}\right)dA_n\;.
\end{equation}
By using equation (\ref{eqn11}), definitions (\ref{eqn10-2}) and
(\ref{eqn10-3}), and the 
transformations (\ref{eqn10-5.5}) and (\ref{eqn10-5.6}), we may write
\begin{eqnarray}
\label{eq43}
dQ&=&\hat v_w dA_w+\hat v_n dA_n=\left(v_w-(1-S_w)v_m\right)dA_w\nonumber\\
&+&\left(v_n+S_wv_m\right)dA_n\;.
\end{eqnarray}
Equating (\ref{eq42}) and (\ref{eq43}) gives equations (\ref{e1}) and (\ref{e2}).

Let us now rewrite the differential $dQ$ as follows,
\begin{eqnarray}
\label{eqn43-1}
dQ&=&d[A_pv]=d[A_p(S_w\hat v_w+(1-S_w)\hat v_n)]\nonumber\\
&=&A_p(\hat v_w -\hat v_n)dS_w+v\ dA_p\;,
\end{eqnarray}
where we have used equation (\ref{eqn1005}).  We now combine this equation with equation (\ref{eq39})
to get
\begin{equation}
\label{eqn43-2}
\frac{dv}{dS_w}=\hat v_w-\hat v_n\;.
\end{equation}
By using equations (\ref{eqn10-5.5}) and (\ref{eqn10-5.6}), we may rewrite this equation in terms of the
seepage velocities, resulting in equation (\ref{eq1003}).

We differentiate $Q$ a second time, which is zero according to the Poincar{\'e} lemma \cite{f63,mtw17}. 
We see that this is indeed so by using equation (\ref{eqn1000}),
\begin{eqnarray}
\label{eqn44}
d^2Q&=&d(\hat v_r dA_r+\hat v_r'A_rd \phi)\nonumber\\
    &=&\hat v_r' \left( d \phi\wedge dA_r \right)+\hat v_r' \left( dA_r\wedge d\phi  \right)\nonumber\\
&=&(\hat v_r'-\hat v_r')d\phi\wedge dA_r=0\;,
\end{eqnarray} 
where the wedge $\wedge$ signifies the antisymmetric exterior product. 

The same equation expressed in the coordinate system $(A_w,A_n)$ gives
\begin{eqnarray}
\label{eq45} 
d^2Q&=&d(\hat v_wdA_w+\hat v_ndA_n)\nonumber\\
&=&\left[\left(\frac{\partial \hat v_n}{\partial A_w}\right)_{A_n}
-\left(\frac{\partial \hat v_w}{\partial A_n}\right)_{A_n}\right]
dA_w\wedge dA_n\nonumber\\
&=&0\;.
\end{eqnarray}
We rewrite the coefficient in terms of the saturation coordinate system,  
\begin{eqnarray}
\label{eqn46}
\left(\frac{\partial \hat v_n}{\partial A_w}\right)_{A_n}
-\left(\frac{\partial \hat v_w}{\partial A_n}\right)_{A_n}&=&\nonumber\\
\frac{d\hat v_n}{dS_w}\left(\frac{\partial S_w}{\partial A_w}\right)_{A_n}-
\frac{d\hat v_w}{dS_w}\left(\frac{\partial S_w}{\partial A_n}\right)_{A_w}&=&\nonumber\\
S_w\frac{d\hat v_w}{dS_w}+(1-S_w)\frac{d\hat v_n}{dS_w}&=&0\;,
\end{eqnarray}
which is nothing but the Gibbs-Duhem equation yet again. Writing this equation in
terms of the seepage velocities gives us equation (\ref{eq1005}). 

We see that the different equations relating the velocities can all be found
from the differential geometric structure of the space spanned by the
areas, with the addition to Euler homogeneity, see Figure \ref{fig3}. All
relations are either a
consequence of writing $dQ$ using different coordinate systems, or a consequence
of the Poincar{\'e} lemma, expressed as $d^2Q=0$.

\section{The co-moving velocity in coordinate-free representation}      
\label{comov}

The co-moving velocity appears in several equations above. 
Common to all of them is that they all are written out in terms of a given coordinate system.
We may write the representation of the co-moving velocity in the saturation coordinate system, equation (\ref{eq1005}), as
\begin{equation}
\label{eqn48}
A_pv_m=\vec A\cdot \frac{d\vec v}{dS_w}\;.
\end{equation}
We may write equation (\ref{eq26}) in polar coordinates as
\begin{equation}
\label{eqn49}
A_r\hat v_m=-\vec A\cdot \frac{d\vec v}{d\phi}\;.
\end{equation} 
Expressed in terms of the seepage velocities, we have that
\begin{equation}
  \label{eq:dQ-seepage}
  dQ  \ = \  v_w dA_{w} + v_{n }d A_n + A_w dv_w + A_n dv_n\;.
\end{equation}

In order to obtain an expression for the co-moving velocity which is independent of
coordinate system, we construct
\begin{eqnarray}
\label{eqn50}
&&A_wdv_w+A_ndv_n=v_m\ \frac{A_ndA_w-A_wdA_n}{A_w+A_n}\nonumber\\
&=&v_m A_pdS_w=-\hat v_m A_r d\phi\;,
\end{eqnarray}
where the second line only reflects equations (\ref{eqn48}) and (\ref{eqn49}).

We write $dQ$ as
\begin{align}
  \label{eqn51}
  dQ=d[\vec A\cdot \hat{\vec v}]  \ =& \  \hat{v}^i dA_i \;,
\end{align}
using equation (\ref{eqn1005}). Here, Einstein summation convention has been
applied, with the index $i$ running over
$\left\{ w,n \right\}$ in the basis $\left(A_w, A_n \right)$.
In terms of the seepage velocities, this becomes
\begin{align}
  \label{eqn52}
  dQ=d[\vec A\cdot \vec v]=   v^{i} dA_{i} + A^{i} d v_{i}   \;.
\end{align}
 Combining equation
(\ref{eqn51}), (\ref{eqn52}) and (\ref{eq27}) gives us
\begin{equation}
\label{eqn53}
\hat{v}_{m}^i dA_i + A^i dv_{i}  = 0
\end{equation}
where $ \hat{v}_m^i$ denotes the components of
$\hat{\vec{v}}_{m}$. This holds in any coordinate system, since it is just
obtained by differentiating a function, namely $\vec{A} \cdot \vec{v}$. We have
therefore obtained an expression for the co-moving velocity which is independent
of the chosen coordinate system.

We can introduce \textit{vector-valued differential forms} \cite{mtw17} to make
a connection between the interpretation in terms of (tangent) vectors used up
until now and the language of differential forms. 
Let $\left\{ \vec{e}_i \right\}$ be an arbitrary basis for vectors on the area space as
we have discussed until now, be it area-, saturation- or polar coordinates.
We now take $\vec{A}$ as an example. We write
\begin{equation}
  \label{eq:A-vector-valued-form}
  A \ = \ \beta^i \vec {e}_i \ \;,
\end{equation}
where $A$ is now regarded as a vector-valued differential form (thus seeing $\vec e_i$ as components
written out in the $\beta^i$ basis, i.e., a reversal of view point), 
the $\beta_i$ are $0$-forms, just functions, and the index $i$
runs over the coordinate indices. We can still treat $A$ as a vector, but the
difference in viewpoints is apparent when we apply the exterior
derivative $d$ to $A$, which gives
\begin{equation}
  \label{eq:derivative-vector-valued-A}
  dA = \sum\displaylimits_{i} (\vec{e}_{i} \otimes d\beta_i) = (d\beta^{i})\vec{e}_i \ \;.
\end{equation}
where we in the second equality apply the summation convention and have
suppressed the tensor product of the forms and basis vectors since we are
working over a vector space, as is standard notation in e.g.\ \cite{mtw17}.
Strictly speaking, the exterior derivative $d$ should be replaced by the
\textit{exterior covariant derivative} in this setting, a generalization of $d$
which is defined on both tangent vectors and forms (there is currently no term
in equation~(\ref{eq:derivative-vector-valued-A}) that reflect the changes in
the basis
$\vec{e}_i$). The terms $A^i dv_{i}$ are then possible to interpret in terms of
\textit{connection forms}. However, this is outside the scope of the current
discussion, and we leave this topic for future work.

If we write out equation (\ref{eq:derivative-vector-valued-A}) in the coordinate
system $\left( A_w, A_n \right)$, we get
\begin{equation}
  \label{eq:dA-area}
  dA \ = \ dA_w \vec{e}_w + dA_n \vec{e}_n \ \;.
\end{equation}
With this formalism in mind, we now return to polar coordinates. Equation
(\ref{eqn53}) may be written in these coordinates by using equation
(\ref{e108-1}) and (\ref{e108-2}) to define the vector valued $0$-forms 
\begin{align}
  \omega_r \ =& \ \cos\phi\  \vec{e}_w+\sin\phi\  \vec{e}_n  \;,\label{eq:wr-form}\\
  \omega_\phi \ =& \ -\sin\phi\ \vec{e}_w+\cos\phi\  \vec{e}_n   \;,\label{eq:wphi-form}
\end{align}
which are equivalent to the corresponding basis vectors in equation
(\ref{e108-1}) and (\ref{e108-2}), but interpreteted as $0$-forms. We then compute
\begin{align}
  d\omega_r \ =& \  \vec{e}_\phi d\phi\;,\label{eqn54}\\
  d\omega_\phi \ =& \ - \omega_rd\phi \;,\label{eqn55}
\end{align}
so that 
\begin{align}
  d A = d( A_r \omega_{r }) \ =& \ dA_r \omega_r+A_rd \omega_r \nonumber \\
  =& \ dA_r \vec{e}_r + A_r \vec{e}_{\phi} d\phi \;, \label{eqn56}
\end{align}
where we in the second equality has used that $\omega_r$ is just another way of
writing the basis vector $\vec{e}_r$ that emphasizes its role as a form. 

We can similarly define a vector valued $1$-forms $d\hat{v}$ and $dv$ from
equation~(\ref{eq19}) and (\ref{eq10}) respectively. We then straightforwardly obtain
\begin{align}
  d\hat{v} \ =& \ d\hat{v}_w \vec{e}_w + d\hat{v}_n \vec{e}_n  \label{eq:hat-v-vector-valued-form} \;, \\
  dv \ =& \ dv_w \vec{e}_w + dv_n \vec{e}_n \label{eq:v-vector-valued-form} \;.
\end{align}
We express $dv$ in polar coordinates as
\begin{eqnarray}
\label{eqn57}
dv&=&v_r' \vec{e}_rd\phi+v_r\vec{e}_\phi d\phi+v_\phi'  \vec{e}_\phi d\phi-v_\phi  \vec{e}_r d\phi\nonumber\\
&=&\left[(v_r'-v_\phi) \vec{e}_r +(v_r+v_\phi') \vec{e}_\phi\right]d\phi\;,
\end{eqnarray}
giving
\begin{equation}
\label{eqn58}
A_r {\hat v}_m d\phi= A_r(v_\phi-v_r')d\phi\;,
\end{equation}
or
\begin{equation}
\label{eqn59}
\hat v_m = v_\phi-v_r'\;.
\end{equation}
By combining (\ref{eq14}), (\ref{eq29}) and (\ref{eqn1003}), we obtain the same equation. 
Equation (\ref{eqn53}) written in the saturation coordinate system yields equation (\ref{eq1005}).

\section{Conclusion and discussion}      
\label{conclusion}
 
The aim of this work has been to formulate the immiscible two-phase flow in porous media problem 
as a {\it geometrical problem\/} in area space.  We did this by
\begin{itemize}
  \item defining the two pore area variables $A_w$ and $A_n$ and consider the space they span,
  \item endow this space with different coordinate systems, $(A_w,A_n)$, $(A_p,S_w)$ and 
        $(A_r,\phi)$, pointing out that expressions that using the polar coordinate system simplifies
        the discussion considerably,  
      \item recognizing the meaning of the volumetric flow rate being a
        degree-$1$ homogeneous function when expressing it in polar coordinates,
        and deriving a number of properties related to the different velocities
        in the problem; the seepage velocities, the thermodynamic velocities and
        the co-moving velocity,
  \item using differential forms to derive relations between the different velocities, and lastly,
  \item formulate an expression for the co-moving velocity which is independent of the coordinates
        on the underlying space. 
\end{itemize}

We remind the reader of the following equations. 
The difference between the thermodynamic and seepage velocities may be expressed by combining equations (\ref{eqn51})
and (\ref{eqn52}),
\begin{equation}
\label{eqn150}
dQ=\hat{v}^{i} dA_{i}= A^{i} d v_{i} + v^{i} dA_{i} \;.
\end{equation}
The co-moving velocity is given by equation (\ref{eq27}),
\begin{equation}
\vec v=\hat{\vec v}+\hat{\vec v}_m\;.\nonumber
\end{equation}

These equations lead us to the three central equations summarizing the central results of this paper. 
They are equation (\ref{eq18}), 
\begin{equation}
\vec v\cdot \vec A=Q\;,
\nonumber
\end{equation}
equation (\ref{eq29-1}), 
\begin{equation}
\hat{\vec v}_m\cdot\vec A=0\;,
\nonumber
\end{equation}
and equation (\ref{eqn53}),
\begin{equation}  
\hat{v}_{m}^i dA_i + A^i dv_{i} = 0
\nonumber
\end{equation} 

It is clear from this discussion that the co-moving velocity, $v_m$, which together with the average seepage velocity
$v$, determines $v_w$ and $v_n$ through the transformations (\ref{e1}) and (\ref{e2}), cannot be found by any
measurement of the volumetric flow rate $Q$ or any expression gotten from it and it alone such as $dQ$.  The 
fundamental question, which remains open is the following: is it possible to measure the co-moving velocity $v_m$
{\it without\/} explicitly measuring $v_w$ and $v_n$?  
\bigskip

The authors thank Dick Bedeaux, Carl Fredrik Berg, Eirik Grude Flekk{\o}y, Magnus Aa.\ Gjennestad, 
Signe Kjelstrup, Marcel Moura, 
Knut J{\o}rgen M{\aa}l{\o}y, Santanu Sinha, Per Arne Slotte and Ole Tors{\ae}ter for 
interesting discussions.  This work was partly supported by the Research Council of Norway 
through its Centres of Excellence funding scheme, project number 262644.


\end{document}